\documentclass[twocolumn,showpacs,preprintnumbers,amsmath,amssymb]{revtex4}

\usepackage{graphicx}
\usepackage{dcolumn}
\usepackage{bm}

\usepackage{textcomp}

\begin{document}

\preprint{APS/123-QED}

\title{Crosstalk Correction in Atomic Force Microscopy}

\author{\'{A}.~Hoffmann}
\author{T.~Jungk}
\author{E.~Soergel}
\email{soergel@uni-bonn.de}

\affiliation{Institute of Physics, University of Bonn,
Wegelerstra\ss e 8, 53115 Bonn, Germany}

\date{\today}

\begin{abstract}
Commercial atomic force microscopes usually use a four-segmented
photodiode to detect the motion of the cantilever via laser beam
deflection. This read-out technique enables to measure bending and
torsion of the cantilever separately. A slight angle between the
orientation of the photodiode and the plane of the readout beam,
however, causes false signals in both readout channels, so-called
crosstalk, that may lead to misinterpretation of the acquired data.
We demonstrate this fault with images recorded in contact mode on
ferroelectric crystals and present an electronic circuit to
compensate for it, thereby enabling crosstalk-free imaging.
\end{abstract}


\maketitle

The atomic force microscope (AFM) has become a standard tool for
determining the surface properties on the nanometer scale not only
in physics but also in all life sciences. This is mainly due to its
high versatility as it can detect various surface properties such as
topography as well as e.g.\ frictional, electrostatic or magnetic
interaction between tip and sample (see e.\,g. \cite{Mey}). This
feature of the AFM is even more attractive since those surface
properties can be detected simultaneously by using an appropriate
setup. Unfortunately an unambiguous separation of the different
read-out channels is not generally assured, leading to crosstalk.
Although commercially available AFM's are generally equipped with a
powerful software for operation and subsequent image processing, a
correction for crosstalk is not provided. In this contribution, we
address the problem of crosstalk between the read-out channels for
bending and torsion of the cantilever.

\begin{figure}
\includegraphics{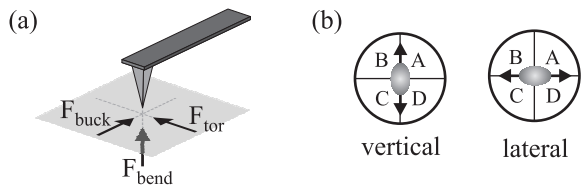}
\caption{\label{fig:hoffmann01}
(a) Forces acting on the tip. $F_{\rm bend}$: bending of the
cantilever due to forces out of plane, $F_{\rm buck}$ and $F_{\rm
tor}$: buckling and torsion of the cantilever due to forces in plane
with the surface.
(b) Readout with the position sensitive detector, left: vertical
signal (bending \& buckling), right: lateral signal (torsion).
}
\end{figure}

Figure~\ref{fig:hoffmann01} shows the notations used. The forces
sensed by the tip can be out of plane (i) and in plane (ii) of the
surface to be investigated. Whereas (i) leads to a bending of the
cantilever, (ii) results either in torsion or in buckling, depending
on the orientation of the force with respect to the axis of the
cantilever. Note that bending and buckling lead to a ''vertical
signal'', i.\,e., the movement of the cantilever is detected as
$(A+B)-(C+D)$ at the position sensitive detector whereas torsion is
seen as a ''lateral signal'' via $(A+D)-(B+C)$.

\begin{figure}
\includegraphics{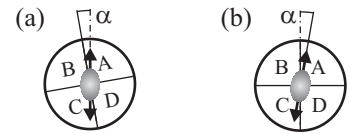}
\caption{\label{fig:hoffmann02}
Misalignment between read-out laser beam and the position sensitive
detector. A rotation of the photodiode (a) affects the readout
likewise a rotation of the plane of the laser beam (b). Both cause
an angular mismatch $\alpha$ that involves false signals due to
crosstalk between vertical and lateral signals.
}
\end{figure}

There are several reasons for crosstalk between the vertical and the
lateral readout channel in AFM: (i) mechanically caused, (ii)
originated by the electronics and (iii) due to a misalignment of the
optical detection system. The first one (i) generally arises when
mechanically hitting an edge at the surface while scanning, thereby
twisting the cantilever. In some AFM's also the elongation of the
tubescanner results in a change of the detection unit, thus, leading
to false signals \cite{Var03,Kwo03}. Finally a mechanical coupling
of the different motions of the cantilever can lead to crosstalk
\cite{Jeo04}. Mechanical crosstalk is in particular important when
investigating samples with a pronounced topography. For reduction a
low scanning speed together with a fast feedback loop is most
appropriate. Note that on smooth sample surfaces mechanically caused
crosstalk does not occur.

Concerning the crosstalk (ii) originating from electronics, a
careful shielding of the signal wires seems most promising. This,
however, can generally be assured by the manufacturer only, the user
having no access to the electronics in the AFM head.

The last type of crosstalk (iii) is generated by a misalignment of
the optical detection unit. The way to adjust the laser beam on the
backside of the cantilever and, in a subsequent step, to center its
reflection on the position sensitive detector (PSD) is most probably
different for every AFM. In addition, to achieve a perfect
alignment, it would be necessary to rotate the PSD thereby avoiding
an angular mismatch $\alpha$ between the axis of the PSD and the
plane of the read-out laser beam (Fig.~\ref{fig:hoffmann02}). The
latter is given by the incoming laser beam and the one reflected
from the cantilever. Although this problem is described in the
literature \cite{Pin02}, a rotation of the PSD is in general not
possible. In case of a misalignment by the angle $\alpha$ the
correct vertical and lateral signals for bending and torsion ($V$
and $L$) are falsified to the measured signals ($V_{\rm m}$ and
$L_{\rm m}$) via the rotation matrix as
\begin{equation}
\left[
  \begin{array}{c}
    V_{\rm m} \\
    L_{\rm m} \\
  \end{array}
\right]
=
\left[
  \begin{array}{c}
    V \\
    L \\
  \end{array}
\right]
\left[
  \begin{array}{cc}
    \cos \alpha & \sin \alpha \\
    -\sin \alpha & \cos\alpha \\
  \end{array}
\right] \; .
\end{equation}

\begin{figure}[ttt]
\includegraphics{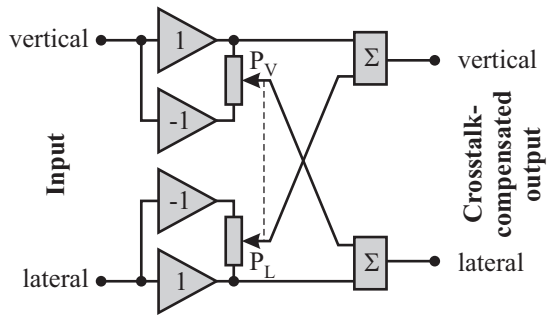}
\caption{\label{fig:hoffmann03}
Schematics of the electronic circuit used for crosstalk
compensation. $P_{\rm V}$ and $P_{\rm L}$: potentiometer for
vertical and lateral crosstalk correction respectively. $\Sigma$:
summing up of the signals.
}
\end{figure}

To correct for this misalignment we realized an electronic circuit
depicted in Fig.~\ref{fig:hoffmann03}, thereby adding separately to
every readout channel a component from the other channel with the
adequate amplitude, adjustable via potentiometers. The
crosstalk-corrected signals $V_{\rm c}$ and $L_{\rm c}$ can thus be
calculated by

\begin{equation}
\left[
  \begin{array}{c}
    V_{\rm c} \\
    L_{\rm c} \\
  \end{array}
\right]
=\left[
  \begin{array}{c}
    V_{\rm m} \\
    L_{\rm m} \\
  \end{array}
\right]
\left[
  \begin{array}{cc}
    1 & -x \\
    x & 1 \\
  \end{array}
\right]
= \frac{1}{\cos \alpha}
\left[
  \begin{array}{c}
    V \\
    L \\
  \end{array}
\right]
\end{equation}
with $x=\tan\alpha$.  The circuit of Fig.~\ref{fig:hoffmann03} was
realized with low-noise precision operational amplifiers (OP27) and
applicable to frequencies $> 1$\,MHz. This is generally enough for
standard AFM applications. Note that the corrected signals become
larger than the real signals. Therefore the calibration of the
system has to be performed after accomplishing the crosstalk
correction or use an adjustable voltage divider at the output.

How to adjust the potentiometers $P_{\rm V}$ and $P_{\rm L}$? In a
first step, the determination of the crosstalk is required. This can
be achieved by retracting the tip from the surface and exciting the
cantilever at its resonance frequency (with help of the piezo used
for non-contact mode operation). The spring constants and
accordingly the resonance frequencies for bending and torsion of a
cantilever are different, thus using the adequate excitation
frequency the cantilever oscillates in its first bending mode only.
In case of a perfect alignment of the optical detection unit $L_{\rm
m}=0$, i.e., no lateral signal is detected. Otherwise the
potentiometer $P_{\rm L}$ has to be adjusted to obtain $L_{\rm
m}=0$. Since both channels suffer the same crosstalk, i.e., the same
rotation $\alpha$, $P_{\rm V}$ has to be set to the same value as
$P_{\rm L}$.  Note that this procedure has to be repeated for every
cantilever, and even for a new laser beam adjustment with the same
cantilever.

\begin{figure}[ttt]
\includegraphics{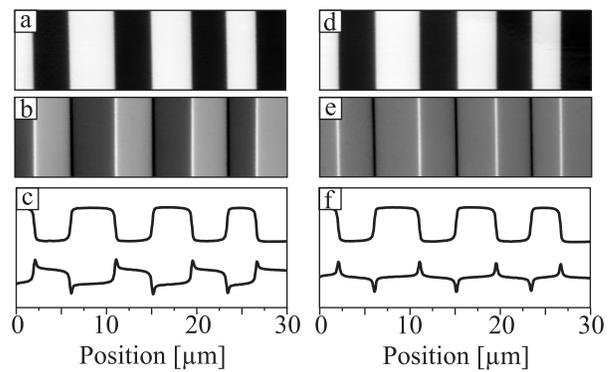}
\caption{\label{fig:hoffmann04}
Piezoresponse force microscopy images of periodically poled lithium
niobate (PPLN). Left side with crosstalk and  right side with
compensated crosstalk ((a), (d): vertical signal, (b), (e): lateral
signal, (c), (f): appropriate scanlines). The cantilever is
orientated parallel to the PPLN stripes.
}
\end{figure}

To give an example of the efficiency of our electronic crosstalk
compensator, we performed measurements on periodically poled lithium
niobate (PPLN) crystals using the AFM in piezoresponse mode
\cite{Jun06b}. Using PPLN as a test sample has the advantage that
the surface deformation at the domain boundaries caused by the
converse piezoelectric effect has a hight of only $< 0.1\,$nm over a
length scale of $\sim 100\,$nm (with $10\,\rm V_{pp}$ applied to the
tip) \cite{Jun06c}. Thus a mechanical crosstalk can be neglected.
The vertical signal is known to be caused by the mechanical
deformation of the sample via the converse piezoelectric effect. The
lateral signal at the domain boundaries originates from the surface
charges leading to electric fields orientated in plane of the
surface \cite{Jun06a}. The left side of Fig.~\ref{fig:hoffmann04}
shows simultaneously recorded deflection (a) and torsion (b) of the
cantilever, without crosstalk correction. In (c) scanlines of these
two images are presented. Obviously also in the lateral channel the
domain faces of PPLN are visible. Because the vertical signal having
a much higher amplitude than the lateral signal, the reciprocal
effect is not seen. When using crosstalk compensation (right side of
Fig.~\ref{fig:hoffmann04}) the lateral signal shows no contrast of
the domain faces but only the boundaries are visible.

A crosstalk compensation as presented above could of course also be
realized by an subsequent software processing of the recorded
images. However, compared to the hardware solution proposed in this
contribution, there are several drawbacks: (i) for the determination
of the correction parameter (the angle $\alpha$), a separate
detection of the vertical and the lateral signal amplitudes ($V_{\rm
m}$ and $L_{\rm m}$) of the excited cantilever is required.
Furthermore, their relative phase relation must be known to identify
the sign of the necessary rotation. These signal parameters,
however, are  not accessible in general. (ii) For crosstalk
compensation via software both images (lateral and vertical) are
necessary since image processing takes only place after recording.
(iii) This implies that a real-time monitoring of the data during
image acquisition is not possible. (iv) Finally, a software based
solution limits the possibilities to record freely chosen input
signals (e.g. the outputs of two lock-in amplifiers as demonstrated
in the above presented example). Note that the drawbacks as
described above could be solved by the manufacturer with a software
compensation during data acquisition and additional hardware
modifications of the control unit.

In this contribution we have demonstrated the effect of a
misalignment of the optical detection unit on the recording of
bending and torsion signals with AFM. We have furthermore proposed
an electronic circuit to compensate for false signals caused by this
type of crosstalk which can be incorporated to every AFM if the
outputs of the position sensitive detector are directly accessible.

\section*{Acknowledgments}
Financial support of the DFG research unit 557 and of the Deutsche
Telekom AG is gratefully acknowledged.

\end{document}